\documentclass[a4paper,12pt,reqno,superscriptaddress,nofootinbib]{revtex4}
\usepackage[centertags]{amsmath}
\usepackage{amsfonts}
\usepackage{amssymb}
\usepackage{amsthm}
\usepackage{newlfont}
\usepackage{stmaryrd}
\usepackage{mathrsfs}
\usepackage{mathtools}
\usepackage{euscript}
\usepackage{graphicx}
\usepackage{enumerate}
\usepackage{todonotes}
\usepackage[normalem]{ulem} 

\usepackage{color}
\usepackage{floatrow}
\usepackage{caption}

\usepackage{tikz}
\usepackage{pgf}
\usetikzlibrary{positioning,fit,calc}
\usetikzlibrary{arrows,automata}
\usepackage{wrapfig}
\usepackage{subfigure}
\usepackage{amscd}
\usepackage{hyperref}

\theoremstyle{plain}

\theoremstyle{definition}

\theoremstyle{remark}

\newcommand{\dd}{ \mathrm{d}}

\newcommand{\bbR}{{\mathbb R}}

\newcommand{\opunit}{\text{1}\kern-0.22em\text{l}}

\DeclareMathAlphabet{\mathpzc}{OT1}{pzc}{m}{it}

\newcommand{\id}{\textrm{d}}

\def\R{\mathbb R}

\let\oldsqrt\sqrt
\def\sqrt{\mathpalette\DHLhksqrt}
\def\DHLhksqrt#1#2{%
\setbox0=\hbox{$#1\oldsqrt{#2\,}$}\dimen0=\ht0
\advance\dimen0-0.2\ht0
\setbox2=\hbox{\vrule height\ht0 depth -\dimen0}%
{\box0\lower0.4pt\box2}}

\begin{document}

\title{Deriving GENERIC from a generalized fluctuation symmetry}

\author{Richard Kraaij}
\affiliation{Fakult\"at f\"ur Mathematik, Ruhr-Universit\"at Bochum}
\author{Alexandre Lazarescu}
\affiliation{Physique et Mat\'eriaux, Universit\'e du Luxembourg}
\author{Christian Maes}
\affiliation{Instituut voor Theoretische Fysica, KU Leuven}
\author{Mark Peletier}
\affiliation{Institute for Complex Molecular Systems, TU Eindhoven}

\begin{abstract}
Much of the structure of macroscopic evolution equations for relaxation to equilibrium can be derived from symmetries in the dynamical fluctuations around the most typical trajectory.  For example, detailed balance as expressed in terms of the Lagrangian for the path-space action leads to gradient zero-cost flow.  We find a new such fluctuation symmetry that implies GENERIC, an extension of gradient flow where a Hamiltonian part is added to the dissipative term in such a way as to retain the free energy as Lyapunov function.
\end{abstract}

\maketitle


\section{History and outline of the paper}
While macroscopic equations describing the return to equilibrium have been conceived and applied even before the atomistic picture of matter was widely accepted, their derivation shows important mathematical and conceptual difficulties.  After all, hydrodynamic and thermodynamic behavior is described autonomously in only a few macroscopic variables and it must be understood how these variables get effectively decoupled from the many microscopic degrees of freedom. Moreover, in modifying the scale of description, the character of the dynamics could drastically change, from a unitary or a Hamiltonian to a dissipative evolution as possibly one of the most remarkable features\footnote{Perhaps the earliest example where that question became manifest is through d'Alembert's paradox (1752) for reconciling, what we call today, Euler's equation with that of Navier-Stokes. As d'Alembert wrote indeed, ``It seems to me that the theory (potential flow), developed in all possible rigor, gives, at least in several cases, a strictly vanishing resistance, a singular paradox which I leave to future Geometers to elucidate.''}.\\

One of the very first and still much studied examples in transiting from microscopic laws to macroscopic behavior is the emergence of the Boltzmann equation for dilute gases.  Ludwig Boltzmann derived the equation in 1872 using a number of dynamical assumptions but its proof and corresponding correct conceptual status were only given a hundred years later \cite{Lanford}.  Oscar Lanford investigated the Boltzmann-Grad limit for a hard sphere gas undergoing Newtonian dynamics and he specified a class of initial conditions under which the Boltzmann equation typically obtains in the macroscopic (kinetic) limit\footnote{A recurrent reaction to that result is the reminder that Lanford's proof only gives the Boltzmann equation for a very short time.  While that is strictly true and for very much understandable technical reasons, such remark appears to us to be similar to emphasizing that Neil Armstrong walked less than 100 meters on the moon.}\cite{bod}.  Boltzmann also showed his famous H-theorem proving in a good sense more than  the second law of thermodynamics by giving a functional, the entropy, that increases along the solution of the Boltzmann equation.  While that proof is often repeated in the form of a simple computation directly working on the time-derivative of the entropy, the reason why it works is much more interesting and was of course captured by Boltzmann's microscopic derivation of irreversibility.  Boltzmann thereby introduced a third scale of description in which the fluctuations are still visible and understood the macroscopic limit as a law of large numbers.  Even when we cannot derive and even when we do not know the precise macroscopic evolution equation, fluctuation theory still tells us that the entropy functional must be increasing in time along that evolution.   The point is that the entropy is the large deviation rate function for macroscopic fluctuations and is therefore \emph{always} increasing along the first-order evolution equation for the macroscopic variable.  That result has been made precise and illustrated in various contexts \cite{dmn2003,villani,jona}.\\ 

A very famous next result was that of Lars Onsager in 1931 \cite{O}, showing that the linearized macroscopic evolution has a symmetry, called reciprocity and appearing in the symmetry of the matrix of the linear response coefficients as a consequence of microscopic reversibility. 
Again fluctuations entered, as reversibility implies that the typical return path to equilibrium can be obtained from the small fluctuations away from  that equilibrium.  That study was continued by Onsager and Machlup in 1953 on the same level of linearized hydrodynamics pioneering there the connection between path-space large deviations and the structure of macroscopic evolution equations \cite{OM}.\\

Linear response theory around equilibrium was formulated systematically in the 1960-70's by various groups with the Green-Kubo and Kubo relations as ultimate results of a first-order perturbation theory. Probably because of the particular quantum style of that time which incited people to start from formal expansions on the level of the Liouville equation, there was little attempt to connect the result with Boltzmann's and Onsager's original line of thinking. The connection with fluctuation theory on path-space was however taken back up again in the derivation of response relations around nonequilibrium, \cite{fdr}.\\
 The mathematical literature on path--wise fluctuations started in the 1970's where large deviation results were obtained for Markov processes and for stochastic perturbations of dynamical systems \cite{DV,FW,fen}.  A non-perturbative path-wise large deviation result within the theory of smooth dynamical systems was the fluctuation theorem of Gallavotti-Cohen for the phase space contraction, \cite{GC,R,verbitski}.   In various subsequent papers that theorem was repeated and connected with a symmetry in the stationary distribution of the time-integrated physical entropy flux for open systems in contact with spatially separated different equilibrium reservoirs, \cite{LS,Kurchan,gibbsian}.  It was realized that the fluctuation symmetry is a path-wise large deviation version of the condition of local detailed balance, \cite{ldb1,ldb2,ldb3} and in that unified way goes together with a greater variety of fluctuation symmetries including the transient regime in the Jarzynski relation, \cite{poincare,crooks}.\\
 
   For all these results one usually starts from more microscopic considerations or from mesoscopic models and moves on to show aspects of the structure of path-wise large deviations.  There is however also the other side, which was in the original goal of Onsager and Machlup, namely to learn  about the structure of the macroscopic evolution equations \emph{directly from} the path-wise large deviations.  That issue has been recently revived by a probabilistic look at so called gradient dynamics.  Our main example and inspiration is \cite{MPR13} which is perhaps still less-known and which we will present in a statistical physics language in Section \ref{grgr}.\\

A substantial part of dissipative relaxational evolution is characterised by what is called \emph{gradient flow}. The broader context here goes back to Josiah Gibbs and the convex geometric way of representing thermodynamic behavior. So called geometric thermodynamics has often been entertained \cite{tis}, where the return to equilibrium is specified geometrically.  Gradient flow is such a characterization, and we can summarize its character by saying that the dynamics proceeds by moving along the gradient of a free energy landscape. That has interesting consequences concerning the relaxation time --- positive Ricci curvature on the free energy landscape is shown to imply a spectral gap --- and for obtaining a geometric view on the relaxation process towards equilibrium \cite{vil}.  It also implies a variational characterization of equilibrium as minimizing the appropriate free energy functional and the entropy-production rate; see the Glansdorff-Prigogine analysis of \cite{GP,zim}.
It has been understood recently how this gradient flow structure is connected with a symmetry in the path-wise large deviations.  In essence, the authors of \cite{MPR13} show that detailed balance, as expressed via the time-antisymmetric part of the Lagrangian for path-wise large deviations, directly leads to the structure of gradient flow.  We repeat the argument in Section \ref{grgr}.   The goal of the present paper is to extend it to GENERIC.\\
The Generalised Equation for Non-Equilibrium Reversible-Irreversible Coupling (GENERIC, \cite{GEN}) corresponds to a class of evolution equations describing return to equilibrium, where besides the dissipative and gradient part in the equation there is also a Hamiltonian part creating a stationary current. Details and examples illustrating our main result can be found in Section \ref{gensec}. 
Our argument exhibits a new fluctuation symmetry on the level of the path-wise large deviations, which leads to GENERIC rather than gradient flow.  That is done by taking the Hamiltonian current as a reference and to study the time-antisymmetric fluctuations \emph{around} it.  The zero-cost flow corresponding to the minimum of the Lagrangian in the path-wise action functional then gives the gradient contribution as added to the Hamiltonian evolution.\\

We start with a section describing what we mean by gradient flow and GENERIC. Besides the general review we also present our first result, which is the characterization of a general form, in equation \eqref{preGENnlin}, of nonlinear relaxation to equilibrium.  In our set-up GENERIC becomes a special case of an evolution which can be characterized as dissipative relaxtion to equilibrium in a moving frame.  In GENERIC the moving frame is the Hamiltonian flow.

\section{Gradient and GENERIC dynamics}

In this section, we introduce the notions of gradient flow and GENERIC, and illustrate them through several examples.

\subsection{Gradient flow}\label{grf}
Gradient flow refers to a certain structure in the relaxation to equilibrium.  The equilibrium itself is characterized in terms of a (set of) macroscopic variable(s), which we denote by $z$ in all abstract generality, but which we will replace within specific examples, depending on the context, by $\rho$ for densities of matter, $x$ for magnetization or $(q,p)$ for a phase space point, {\it etc}. The evolution of the variable $z$ is determined by the current $j_z$ through an equation of the form 
\begin{equation}\label{zj}
\dot z = D j_z
\end{equation}
where the operator $D$ can be minus the divergence, the identity operator, or a stoichiometry matrix, depending on the type of dynamics being considered (respectively: type-B, typically for a density; type-A, like for magnetization; chemical reactive systems).\\

Quite abstractly an evolution has traditionally been called a gradient flow if the displacement of the dynamical variable $z$ can be written in the form
\begin{equation}\label{grade}
\dot z = M\,\id S,\qquad\;\; M = D\, X\, D^\dagger,\qquad\;\; j_z = X\,D^\dagger \,\id S
\end{equation}
for a symmetric positive semi-definite operator $X$ and the adjoint $D^\dagger$ defined by the relation $a\cdot Db=b\cdot D^\dagger a$. Here $\id S$ stands for the derivative of a state function $S(z)$ with respect to $z$ (being a functional derivative if $z=\rho$). $S$ might be an entropy, minus a free energy or some other thermodynamic potential depending on the context, {\it etc}.
Examining the time-derivative of $S$, we find 
\begin{equation}\label{1mo}
\dot S =\id S \cdot \dot z= \id S \cdot M\, \id S \geq 0
\end{equation}
so that $S$ is always increasing and the equilibrium state is obtained by maximizing $S$.
Mathematically it makes sense to think of $z$ as a point of a Riemannian manifold, moving by steepest descent of the functional $-S$. The descent is measured in a metric provided by the operator $X$, which is in general related to the physical mobility.\\

In \cite{MPR13}, the notion of gradient flow was extended from \eqref{grade} to nonlinear evolutions, which appear for instance when considering reactive processes or jump processes on a discrete state space, with Poissonian statistics \cite{mpel}. The nonlinear gradient flow equation, which still describes relaxation towards equilibrium but with a nonlinear operator `$X$', is obtained by replacing \eqref{grade} with
 \begin{equation}\label{gger}
\dot z =D~\partial \psi^\star(D^\dagger \id S/2;z) 
 \end{equation}
for a convex functional $f \mapsto \psi^*(f;z)$ where $f$ stands for a thermodynamic force. An essential requirement here is that 
\begin{equation}
\label{cond:psi*-min-at-zero}
\psi^*(f;z)\geq \psi^*(0;z)=0,
\end{equation}
and that $\psi^*$ is symmetric in $\pm f$; the nonlinear gradient-flow structure~\eqref{gger} generalizes Onsager's reciprocity relations to nonlinear mobility operators~\cite{pons}\footnote{This can be understood as follows: the Onsager reciprocity property $X=X^T$ is equivalent to the property that $Xf = \partial \psi^\star_X(f)/2$ for  $\psi^\star_X(f) = f\cdot Xf$. In~\cite{pons} the authors derive the variational form $\partial \psi^*(f)$ from the same large-deviation considerations that are implicitly present in Onsager's original work~\cite{O}, without imposing Onsager's close-to-equilibrium condition.}. As in \eqref{1mo}, we can write
\begin{equation}\label{ly}
\dot S =\id S \cdot \dot z=2~D^\dagger\id S/2 \cdot \partial \psi^*( D^\dagger \id S/2) \stackrel{\eqref{cond:psi*-min-at-zero}}\geq 0,
\end{equation}
which gives again the monotonicity of $S(z(t))$ in time. Note that we will always consider the derivative $\partial$ to act only on the first variable, treating $z$ as a parameter and sometimes omitting it from our notations. Note also that the linear  case \eqref{grade} can be trivially recovered in these terms by considering for \eqref{gger} the special case where $\psi^\star(f) = f\cdot Xf$.
\\
 
To be more specific, we examine a few examples: a diffusion, a jump process, and a nonlinear process obtained as a rescaling of a jump process.

\subsubsection{Example 1: type-B dynamics}\label{ex1} 

The most standard example of gradient flow is given by a conservative relaxational dynamics (type-B) in the form of a continuity equation for a scalar density $\rho(r,t)$ defined for every position in a  fixed volume, $r\in \Lambda\subset \R^3$, and with fixed boundary conditions $\rho(r,t) = \bar{\rho}, r\in \partial \Lambda$:
\begin{equation}\label{cequ}
\dot\rho = - \nabla \cdot j_\rho 
\end{equation}
where $D=-\nabla\cdot$ is (minus) the divergence with respect to $r$ so that $D^\dagger=\nabla$ is the gradient. The current  is given in terms of a density-dependent mobility $\chi$ (being a $3\times 3$ symmetric non-negative matrix) and the thermodynamic force $F$, as
\begin{equation}\label{lmu}
j_\rho( r)
= \chi\left(\rho( r)\right)\, F( r),\quad F( r)
= - \nabla \,\mu( r),\quad \mu( r) = \frac{\delta{\cal F}[\rho]}{\delta \rho( r)}
\end{equation}
for  local chemical potential $\mu$  expressed as a variational derivative from free energy functional ${\cal F}[\rho]$ up to a constant.  Such a dynamics \eqref{cequ}--\eqref{lmu} gives a description of hydrodynamic relaxation to equilibrium, or, ignoring space-time rescaling, it can be part of a description in  
dynamical density functional theory \cite{rev,rev2} to describe an inhomogeneous fluid. Given \eqref{lmu} we can rewrite \eqref{cequ} as
\begin{equation}\label{he}
\dot{\rho}(r,t)=\nabla\cdot\left(\chi(\rho(r,t)))\nabla \left(\frac{\delta{\cal F}[\rho]}{\delta \rho( r)}\right)\right)
\end{equation}
to recognize the gradient flow structure \eqref{grade} with $M=-\nabla\cdot\chi\nabla$ and $S=-{\cal F}$. The monotonicity of the entropy, or equivalently of the free energy, is
\[ 
\frac{\id}{\id t}{\cal F}[\rho] = -\int \mu(r) \,\nabla \cdot j\left(\rho( r)\right)\id r = -\int \nabla \mu(r) \cdot \chi\left(\rho( r)\right)\nabla \mu(r)\id r  \leq 0
\] 
where the second equality uses partial integration with vanishing boundary term, and the final inequality is obtained by the non-negativity and symmetry of the mobility matrix $X=\chi$ as in \eqref{1mo}.

For example, on the unit interval $\Lambda=[0,1]$ with $\rho(0,t) =  \rho(1,t) = \bar{\rho}$ we can take the grand potential at temperature $T$,
\[ 
{\cal F}[\rho] = k_BT \int_0^1 \left( \rho(r)\,\log \frac{\rho(r)}{\bar{\rho}}-\rho(r)+ \bar{\rho} \right) \id r,\quad \text{ for which } \mu(r) =k_BT \log \frac{\rho(r)}{\bar{\rho}}
\] 
and \eqref{he} is then the linear diffusion equation for the mobility $\chi = \rho$.\\

\subsubsection{Example 2: Markov jump process}\label{ex2} 

We next look at  a detailed-balanced Markov jump process as an example of a type-A relaxational dynamics on a discrete state space.  The macroscopic equation is here the Master equation
\begin{equation}\label{me}
\dot\rho(c)=\sum\limits_{c'}k(c',c)\rho(c')-k(c,c')\rho(c)
\end{equation}
for the time-dependent density  $\rho(c) = \rho(c;t)$ on configurations $c$, with transition rate $k(c,c')$ for the jump $c \rightarrow c'$.  Detailed balance at inverse temperature $\beta$ with respect to a potential $V(c)$ is enforced by requiring $k(c,c')=\varphi(c',c)\mathrm{e}^{\frac{\beta}{2}(V(c)-V(c'))}$, where $\varphi(c',c)=\varphi(c,c')\geq 0$ are symmetric activity parameters for the transitions $c\leftrightarrow c'$.\\
The operator $D$ of \eqref{grade} is here minus the discrete divergence acting on antisymmetric functions $b$,  and $D^\dagger$ acts as a gradient on functions $a$ of configurations $c$:
\begin{equation}\label{dive}
(Db)(c) = -\sum_{c':\varphi(c',c)\neq 0}b(c,c'),\quad a\cdot D b=\sum\limits_{c\sim c'}a(c)b(c',c),\quad (D^\dagger a)(c,c') = a(c') - a(c)
\end{equation}
where the sum in the middle is over all $(c,c')$ where $\varphi(c,c')\neq 0$.
The current in \eqref{me} over $c\rightarrow c'$ is
\[
j(c,c')=k(c,c')\rho(c) - k(c',c)\rho(c')
\]
but can also be written as
\begin{equation}
j(c,c')=\varphi(c,c')\sqrt{\rho(c)\rho(c')}\left(\mathrm{e}^{-\frac{1}{2}D^\dagger[\log \rho +\beta V](c,c')}-\mathrm{e}^{\frac{1}{2}D^\dagger[\log \rho +\beta V](c,c')} \right)
\end{equation}
We therefore take
\begin{equation}\label{Sjump}
S[\rho]= -\sum\limits_c \rho(c)\left[\log\rho(c)+\beta V(c)\right]
\end{equation}
to express the Master equation \eqref{me} in its nonlinear gradient flow form \eqref{gger}, with
\begin{equation}
\psi^\star(f;\rho)= 2\sum\limits_{c,c'} \varphi(c,c')\,\sqrt{\rho(c)\rho(c')}\,\left(\cosh  f_{c,c'}-1\right)
\end{equation}
as function of a force $f_{c,c'}$, function of $(c,c')$.

\subsubsection{Example 3: type-A dynamics}\label{ex3}

As a simple example of a non-conservative relaxational dynamics (so called type-A) we can take a scaling limit of a simple Markov jump process: the
Ehrenfest model which is equivalent to the kinetic Ising model with mean field interaction.\\
Consider $N$ spins $\sigma_k=\pm 1$, and their mean $x(\sigma)=\frac{1}{N}\sum\limits_k \sigma_k$. The spins interact at inverse temperature $\beta$ via a potential $NV(x(\sigma))$ which depends only on their mean, so that it is trivial here to obtain an autonomous evolution of $x$ in the limit of large $N$.

The system evolves by flipping individual spins with  generator given by
\[ 
A_N f(\sigma)=\sum\limits_k e^{-\frac{N\beta}{2} \left(V(x(\sigma^{(k)})-V(x(\sigma))\right)}\,[f(\sigma^{(k)})-f(\sigma)]
\] 
where $\sigma^{(k)}$ is obtained by flipping spin $k$ in $\sigma$.  This process satisfies the condition of detailed balance. 
It induces a Markov process on the reduced variable $x(\sigma)$ for any $N$, given by
\begin{align}\label{an}
A_N f(x) =& N\frac{1-x}{2}e^{-\frac{N\beta}{2} \big(V(x+2/N)-V(x)\big)}\,[f(x+2/N)-f(x)]\nonumber\\
&+N\frac{1+x}{2}e^{-\frac{N\beta}{2} \big(V(x-2/N)-V(x)\big)}[f(x-2/N)-f(x)]
\end{align}
with equilibrium distribution $\nu_N(x)\propto \binom{N}{N\frac{x+1}{2}}e^{-N\beta V(x(\sigma))}$.  In the large $N$ limit we obtain a deterministic limiting process, 
\begin{equation}
\dot f(x) = \left((1-x)e^{-\beta V'(x)}-(1+x)e^{\beta V'(x)}\right)f'(x)
\end{equation}
for smooth functions $f$ on $[-1,+1]$. The limiting evolution for the magnetization $x$ thus verifies
\begin{equation}
\dot x = (1-x)e^{-\beta V'(x)}-(1+x)e^{\beta V'(x)}
\end{equation}
which can be rewritten as
\begin{equation}\label{graf}
\dot x=2\sqrt{1-x^2}\,\sinh\left(-\beta w'(x)\right)
\end{equation}
with $w(x)= V(x)-\beta^{-1}\eta(x)$ where $\eta$ is a mixing entropy 
\begin{equation}
\eta(x)= -\,\left[\frac{1+x}{2}\log\left(\frac{1+x}{2}\right) + \frac{1-x}{2}\log\left(\frac{1-x}{2}\right)\right]
\end{equation}
The prefactor $\chi(x)= 2\sqrt{1-x^2}\geq 0$ can be interpreted as a susceptibility and $w$ as a free energy, which we can check to be decreasing in time:
\begin{equation}
\dot w(x) = w'\,\dot x= w'\,\chi(x)\,\sinh(-\beta w'(x))\leq 0
\end{equation}

We indeed recognize for \eqref{graf} the nonlinear gradient flow structure \eqref{gger}, with entropy $S=-\beta w$, and 
\begin{equation}
 \psi^*(f;x) = \sqrt{1-x^2}\,\left(\cosh 2f(x) - 1\right)
\end{equation}
The operators $D=D^\dagger$ of \eqref{gger} are the identity.

\subsection{GENERIC and pre-GENERIC}\label{GENpreGEN}

GENERIC (the General Equation for Non-Equilibrium Reversible-Irreversible Coupling \cite{GEN}), in its original form, is an additive combination of a Hamiltonian and a gradient flow. Symbolically, it extends \eqref{grade} to
\begin{equation}\label{gene}
\dot z = A \, \dd E + M\, \dd S
\end{equation}
where   $E=E(z)$ and $S=S(z)$ are interpreted as energy and entropy functionals, with  $\dd E$ and $\dd S$ the appropriate derivatives.  $A=A(z)$ is an antisymmetric operator while $M=M(z)$ is a symmetric, non-negative definite operator.  Finally, $E,S,M,A$ are assumed to satisfy the orthogonality conditions
	\begin{equation} \label{eqn:orthogonality}
	A \,\dd S = 0, \qquad M\, \dd E = 0
	\end{equation}
As a consequence, along the flow,
\begin{align*}
\dot S & = \dd S \cdot A \,\dd E + \dd S \cdot M\,\dd S = \dd S \cdot M\,\dd S \geq 0\\
\dot E & = \dd E \cdot A\, \dd E +\dd E \cdot M \,\dd S = 0
\end{align*}
In other terms, GENERIC characterizes a dynamics of return to equilibrium consisting of two orthogonal parts, the first being dissipative (the gradient flow along the entropy $S$), whereas the second is Hamiltonian and conserves the energy $E$. 

We consider here a less constrained version of those equations, which can be seen as a precursor to GENERIC, which we call \emph{pre}-GENERIC. It consists in renouncing to the existence of a conserved energy $E$, but instead considering a general flow ${\cal D} J$ instead of $A\,\dd E$, 
\begin{equation}\label{preg}
\dot z ={\cal D} J + M\, \dd S, \qquad M = DXD^\dagger
\end{equation}
with the operator ${\cal D}$  acting on $J$, a given function of $z$, similar to \eqref{zj}, and
where the second part is the gradient flow of equation \eqref{grade}.
The first constraint in \eqref{eqn:orthogonality} can then be replaced by the less constraining orthogonality condition 
\begin{equation}\label{ort}
{\cal D}J\cdot \dd S=0
\end{equation} leading as before to the monotonicity of $S$.
The second constraint in \eqref{eqn:orthogonality}, as well as the conservation resulting from it, are removed as unnecessary for the structure.\\
Like in the case of gradient flow, (pre-)GENERIC can and in fact should be extended to nonlinear flows (see also~\cite{Mielke11a}) as
\begin{equation}\label{preGENnlin}
\dot z ={\cal D} J +D~\partial \psi^*(D^\dagger \id S/2;z)
\end{equation}
The monotonicity of the entropy follows as in \eqref{ly},
\begin{equation}\label{mon}
\dot S  = \dd S \cdot {\cal D} J + 2~\id S/2 \cdot D \partial \psi^*( D^\dagger \id S/2) \geq 0
\end{equation}
Recognizing the structure \eqref{preGENnlin}  as the typical  behavior unifying various types of relaxation to equilibrium is one of the main results of this paper.\\
The appeal of pre-GENERIC is twofold: first, as we will see, it appears in examples in a more natural and physical way than full GENERIC, which often requires adding extra variables to the models; secondly, as shown in \cite{other}, it turns out that the aforementioned extension is systematically possible, meaning that pre-GENERIC formally implies GENERIC.  We are therefore mostly interested here in working with pre-GENERIC. In order to get the full GENERIC structure, we have to add an auxiliary scalar variable $E$ to the system, in order to fix the two conditions that were not satisfied above. We refer to \cite{DPZ13} for one systematic way to do that.\\

As before we now examine a few illustrative examples.

\subsubsection{Example 4: underdamped diffusions}\label{ex4}
As shown in \cite{DPZ13}, the underdamped Vlasov-Fokker-Planck equation naturally gives rise to GENERIC and hence to a triple $(J,M,S)$. It provides us with an opportunity to illustrate the formal notation of \eqref{preg}.\\

The variable $z$ is, as in Example \ref{ex1} above, a time-dependent density $\rho(q,p;t)$, but in its underdamped version, depending on positions $q\in \R^d$ and momenta $p\in \R^d$.  The Vlasov-Fokker-Planck equation is given by
\begin{equation}\label{vfp}
\dot \rho = -\nabla_q\cdot\rho\frac{p}{m} + \nabla_p\cdot\rho\left(\nabla_qV + \nabla_q(\Phi\star\rho)+\gamma\frac{p}{m}\right) + \gamma \,\beta^{-1}\,\Delta_p\rho
\end{equation}
for damping coefficient $\gamma$, mass $m$ and inverse temperature $\beta = (k_BT)^{-1}$.
The convolution is defined as
\[ 
\Phi\star\rho\;(q) = \int_{\R^{2d}} \Phi(q-q')\,\rho(q',p')\,\id q'\,\id p'
\] 
That equation arises as a mean-field limit of underdamped diffusions interacting through a potential $\Phi$, with friction $\gamma$ and at inverse temperature $\beta$ \cite{bouch}. The Kramers equation describing the evolution of a probability density for underdamped Markov diffusions is recovered for $\Phi=0$.\\

In order to unravel the structure of \eqref{vfp}, we rewrite it in matrix form, separating the $q$ and $p$ directions. For instance, the operator $D = {\cal D}$ is given by minus the divergence
\begin{equation}
D=-\nabla=-\begin{bmatrix}~\nabla_q~&~ \nabla_p~\end{bmatrix}
\end{equation}
Moreover, this being an equilibrium system, we can define the Gibbs entropy
\begin{equation}
{\cal S}[\rho] = -k_B \int_{\R^{2d}}\rho\log\rho\,\id q\id p
\end{equation}
and the energy (including kinetic energy, potential self-energy and interaction energy)
\begin{equation}
{\cal H}[\rho] = \int_{\R^{2d}}\left(\frac{p^2}{2m} + V(q) + \frac 1{2}(\psi\star\rho)(q)\right)\,\rho\,\id q\id p
\end{equation}
and combine them into the free energy functional ${\cal F}[\rho]={\cal H}[\rho]- T\, {\cal S}[\rho]$.

The equation \eqref{vfp} can be then written as
\begin{equation}\label{grafi}
\dot \rho = -\nabla\cdot \rho K\,\nabla \frac{\delta{\cal H}[\rho]}{\delta \rho(r)} +\nabla\cdot\chi \nabla\frac{\delta{\cal F}[\rho]}{\delta \rho(r)}
\end{equation}
with $2d\times 2d$ matrices
\begin{equation}\label{mat}
K =\begin{bmatrix} ~~0~& ~1~ \\ -1~& ~0~ \end{bmatrix}~~~~~\mathrm{and}~~~~~\chi=\rho\gamma\begin{bmatrix} ~0~&~0~ \\~0~& ~1~ \end{bmatrix}
\end{equation}

The system evolution \eqref{grafi} (and thus also \eqref{vfp}) has the pre-GENERIC structure \eqref{preg}, with  $S=-{\cal F}, M=-\nabla \cdot \chi \nabla $, and with $J=\rho K\,\nabla \frac{\delta{\cal H}[\rho]}{\delta \rho(r)}$. Indeed, as required for \eqref{mon} we have the orthogonality
\begin{align}
J\cdot D^\dagger\id S&= \int_{\R^{2d}} \frac{\delta{\cal H}[\rho]}{\delta \rho(r)}\nabla\cdot\rho K \nabla \frac{\delta{\cal F}[\rho]}{\delta \rho(r)} \,\id q\id \nonumber\\
&=\int_{\R^{2d}} \frac{\delta{\cal H}[\rho]}{\delta \rho(r)}\nabla\cdot\rho K \nabla \frac{\delta{\cal H}[\rho]}{\delta \rho(r)} \,\id q\id p-\beta^{-1}\int_{\R^{2d}} \frac{\delta{\cal H}[\rho]}{\delta \rho(r)}\nabla\cdot\rho K \nabla \frac{\delta{\cal S}[\rho]}{\delta \rho(r)} \,\id q\id p\label{40}\\
&=0\nonumber
\end{align}
because of the antisymmetry of $K$ for the first term, and the fact that $\nabla\cdot\rho K \nabla \frac{\delta{\cal S}[\rho]}{\delta \rho(r)} =(\nabla_q\nabla_p-\nabla_p\nabla_q)\rho=0$ for the second term in \eqref{40}.

\subsubsection{Example 5: Andersen thermostat}\label{ex5}

A good example of a nonlinear GENERIC system is the so-called Andersen thermostat \cite{and}, often used for numerical simulations of Hamiltonian systems subject to thermal fluctuations. In this model, independent massive particles with positions $q$ and momenta $p$, of mass $m$ and subject to a potential $V(q)$, are perturbed in their Hamiltonian motion by having their momentum randomized according to their natural Maxwellian distribution at constant rate $\varphi\geq 0$. The evolution of the density of particles $\rho$ is then given by
\begin{equation}\label{rhodotAnderson}
\dot \rho(q,p) = -\nabla\cdot\rho(q,p)  K\,\nabla \frac{\delta{\cal H}[\rho]}{\delta \rho(q,p)}+ \varphi\,\frac{e^{-\frac{p^2}{2m}}}{\sqrt{2\pi m}}\int\id p'~\rho(q,p')- \varphi\,\rho(q,p)
\end{equation}
which is a combination of a Hamiltonian flow and a jump process on $p$ for inverse temperature $\beta=1$.
The Hamiltonian is given by
\[ 
{\cal H}[\rho] = \int_{\R^{2d}}\left(\frac{p^2}{2m} + V(q)\right)\,\rho(q,p)\,\id q\id p
\] 
and the symplectic  matrix $K$ is the same as in \eqref{mat}. We can also use the same entropy $\cal S$ and free energy $\cal F$ as there.\\
For the jump part we imagine  replacing all sums by integrals in Section \ref{ex2} getting a nonlinear gradient flow \eqref{preGENnlin} with $S=-\cal F$ and
\begin{equation}
\psi^\star(f;\rho)=\frac{2\varphi}{\sqrt{2\pi m}}\,\int e^{-\frac{p^2+p'^2}{4m}}\,\sqrt{\rho(q,p)\rho(q,p')}\left(\cosh f - 1 \right) \,\id p \id p'
\end{equation}
The corresponding negative divergence  $D$  and gradient $D^\dagger$  are as in \eqref{dive},
\[ 
D b\,(p)=-\int b(p,p')\, \id p' ,\qquad D^\dagger a(p,p') = a(p') -a(p)
\] 
That allows us to write \eqref{rhodotAnderson} as 
\begin{equation}
\dot \rho + \nabla \cdot J = D~\partial \psi^*(D^\dagger \id S/2;z)
\end{equation}
which is of the pre-GENERIC form \eqref{preGENnlin} with $\cal D = -\nabla\cdot$, and $J=\rho K\,\nabla \frac{\delta{\cal H}[\rho]}{\delta \rho(r)}$ and the orthogonality \eqref{ort} as before in \eqref{40}. 

This example can be easily extended to more general jumps and even to a combination of diffusions and jumps, as long as the rates satisfy detailed balance with respect to $\cal H$. This type of system was considered in \cite{massi} and interpreted as being in equilibrium with work-producing reservoirs, where it is necessary to factor out that work in order to obtain the physical entropy production of the system. As we see here, that is equivalent to writing the fluctuation symmetry around the average flux $J$ present in the system.

\subsubsection{Example 6: nonlinear friction}\label{ex6}

We continue with the jump part for $(q,p)\in \bbR^2$ but with transitions $p\rightarrow p\pm N^{-1}$ with appropriately rescaled rates, by analogy with the Ehrenfest model in Section \ref{ex3}. The backward generator $A_N$ on any smooth function $f$ is given by
\begin{align}
A_N f(q,p)=&\left(\nabla f(q,p)\right)\cdot K\,\nabla \frac{\delta{\cal H}[\rho]}{\delta \rho(r)}\nonumber\\
&+N \varphi~e^{-N\frac{(p+N^{-1})^2-p^2}{4m}}[f(q,p+N^{-1})-f(q,p)]\nonumber\\
&+N \varphi~e^{-N\frac{(p-N^{-1})^2-p^2}{4m}}[f(q,p-N^{-1})-f(q,p)]\label{BackGenE6}
\end{align}
Under the $N\rightarrow\infty$ limit, that becomes
\[ 
A f(q,p)=\left(\nabla f(q,p)\right)\cdot K\,\nabla \frac{\delta{\cal H}[\rho]}{\delta \rho(r)}- 2\varphi~\sinh\left(\frac{p}{2m}\right)\nabla_p f(q,p)
\] 
which corresponds to the equations of motion
\begin{align}
\dot{q}&= \frac{p}{m}\nonumber\\
\dot{p}&=-\nabla_q V(q)-2\varphi~\sinh\left(\frac{p}{2m}\right)\label{nlf}
\end{align}
which shows nonlinear friction.\\
Defining the energy $h(q,p)=\frac{p^2}{2m}+V(q)$, we can rewrite it as
\begin{equation}
\begin{bmatrix} \dot q \\ \dot p \end{bmatrix}=K\nabla h+ \partial\psi^\star\left(-\frac{1}{2}\nabla h\right)
\end{equation}
with
\begin{equation}
\psi^\star(f_q,f_p;q,p)=2\varphi\;\left(\cosh f_p-1\right)
\end{equation}
which is of pre-GENERIC form \eqref{preGENnlin} with $D=\cal D= 1$, $J=K\nabla h$ and $S=-h$. The orthogonality \eqref{ort} is verified as
\begin{equation}
J\cdot D^\dagger\id S=-(K\nabla h)\cdot \nabla h =0
\end{equation}
by the antisymmetry of $K$.\\


\section{Large deviations approach to equilibrium flows}

We have studied \eqref{he}, \eqref{me}, and \eqref{graf} as examples of gradient flow, and \eqref{vfp}, \eqref{rhodotAnderson}, and \eqref{nlf} as examples of pre-GENERIC.  Now comes the moment to connect those structures with symmetries in the dynamical fluctuations.

Dynamical ensembles give the probabilities of trajectories for particle systems \cite{fen,com,poincare}.  Those probabilities can arise from random initial conditions and/or coarse graining where the integrated degrees of freedom have been assigned a statistical distribution.  For the present paper we place ourselves at a mesoscopic level, looking at the fluctuation dynamics of macroscopic degrees of freedom in a system with a large parameter $N$ (usually size, number of elements, inverse temperature or number of copies), as could be obtained via the theory of large deviations \cite{FW,fen,lag1,lag2,com,jona}.  The main objects of interest are then the Lagrangian $L$ and the Hamiltonian $H$.\\ 
We consider in a time-interval $[0,t]$ all possible evolutions of the variable $z$ and of a current variable $j$ which are compatible with the microscopic system; we thus have a flow of state  and current variables $(z(s),j(s)), s\in [0,t])$ verifying, at each moment, the equation 
\begin{equation}\label{str}
\dot{z}(s) - \cal D J(z(s)) = D j(s)
\end{equation}
for a given fixed current $J$ and operators $\cal D$ and $D$ as we had them before in the examples. In many interesting cases however $\cal D = D$, and we can then in fact stick to the constraint 
\begin{equation}
\label{stud}
\dot{z}(s) = D j(s)
\end{equation}
where the  $J$ of \eqref{str} is possibly to be part of the $j$.\\
  There are many possible trajectories all satisfying the constraint \eqref{str} or \eqref{stud}, and we investigate the structure of their probability.
At time zero we start in equilibrium so that the probabilities are given by an entropy functional $S\left(z(0)\right)$, while for later times the probability will also involve the currents $j$.\\
The Lagrangian $L(j;z)\geq 0$ of the system governs the path probabilities of the macroscopic variable, via
\begin{equation}\label{prob}
\mathrm{Prob}[(z(s),j(s)), s\in [0,t]] \simeq e^{N S(z(0))}\,e^{-N\int_0^t\id s\, L(j(s);z(s))},\qquad N\uparrow +\infty
\end{equation}
which is to say that $L$ determines the plausibility of the various possible trajectories.  Note that all fluctuations are exponentially damped in $N$ with respect to the zero cost flow $j_z$ where $L(j_z;z)=0$. Properties that can be assumed from the outset are that $L(j,z)\geq 0$, is convex in $j$ for all $z$, and that $L=0$ induces a unique evolution equation, the so called zero-cost flow. Finding the differential equation for the typical macroscopic trajectory is thus equivalent to finding for any  given $z$ the solution $j_z$ of $L(j_z;z)=0$, and then to substitute that solution into \eqref{str} or \eqref{stud} to get an autonomous equation for $z(t)$.  The question we ask here more specifically is 

\begin{quote}
Can we identify the $\psi^*$  in \eqref{preGENnlin} from properties of the Lagrangian $L$, for $S$ given in \eqref{prob}?
\end{quote}
In particular, can we see how symmetries of the Lagrangian can naturally lead to gradient flow and to pre-GENERIC?\\

Note also that from the Lagrangian $L$ we  define the Hamiltonian $H$ through a Legendre transform:
\begin{equation}\label{defH}
H(f;z) = \sup_j\left[j\cdot f - L(j;z)\right],
\end{equation}
where $f$, the variable dual to the current $j$, is a thermodynamic force. By construction, $L(j_z;z)=0$ is equivalent with $H$ verifying
\begin{equation}\label{pH2}
j_z =\partial H(0;z)
\end{equation}
Symmetries of the Lagrangian will be translated in properties of $H$.

\subsection{From detailed balance to gradient flow}\label{grgr}

We first address the question above in the context of detailed balance  (time-reversibility).
Here we assume the constraint \eqref{stud} for some operator $D$ as appears in \eqref{zj}--\eqref{gger} (or, $J=0$ in \eqref{str}).  We will see how under detailed balance conditions the time-symmetric part of the Lagrangian, $L(-j;z) + L(j;z)$, determines the zero-cost flow $j_z$ for given entropy $S$, 
and in such a way that the autonomous evolution is a gradient flow with respect to the entropy $S$, as in \eqref{gger} or as in the examples of Section \ref{grf}.\\

By detailed balance (on the mesoscopic level of \eqref{prob} and for a variable $z$ which is even under kinematic time-reversal) we mean that the probability of any trajectory, including the initial condition distributed with respect to the entropy $S$, is equal to that of its time-reversal.  That is the condition that
\begin{equation}\label{ref}
\int_0^t\id s\, \left[L\left(-j(s);z(s)\right) - L\left(j(s);z(s)\right)\right] = S\left(z(t)\right) - S\left(z(0)\right)
\end{equation}
which, if true for all times $t$, is nothing else than requiring $L(-j;z) - L(j;z) = \dot S$, or that
\begin{equation}
\label{detb}
L(-j;z) - L(j;z) =  j\cdot D^\dagger \id S(z)
\end{equation}
Similarly, under detailed balance \eqref{detb} and from \eqref{defH} we have 
\begin{equation}\label{dbH}
  H(f;z)=H(-f-D^\dagger \id S;z)
\end{equation}
Clearly now, at zero-cost flow $j_z$ where $L(j_z;z)=0$, the time-symmetric part must equal  the antisymmetric part of the Lagrangian, which under \eqref{detb} yields
\begin{equation}
L(-j_z,z) + L(j_z,z) =  j_z\cdot D^\dagger \id S(z) 
\end{equation}
which is an equation for $j_z$.  In order to solve it, we will decompose the time-symmetric part $L(-j;z) + L(j;z)$, as in the left-hand side, in a pair of convex conjugates.\\

Define
\begin{eqnarray}
\psi(j;z) &=& \frac{1}{2}\bigl[L(-j;z) + L(j;z)\bigr]-L(0;z) \nonumber\\
 &=& \frac{1}{2}\,j\cdot D^\dagger \id S(z) + L(j;z) -L(0;z)  \label{Lpsi}
\end{eqnarray}
where the second line uses \eqref{detb} and implies that $\psi(j;z)$ is convex in $j$.
From the first line we see that $\psi$ is symmetric in $\pm j$ and vanishes at $j=0$.  It is therefore also positive, $\psi(j;z)\geq 0$. 
The Legendre transform of $\psi$ is
\begin{equation}
\psi^\star(f;z) = \sup_j\left[j\cdot f - \psi(j;z)\right]
\end{equation}
and we will show that it is the same function as appears in \eqref{gger}.  Note indeed that by replacing here $\psi$ via \eqref{Lpsi} we find a convex
\begin{equation}\label{Hpsistar1}
\psi^\star(f;z) = H(f-D^\dagger \id S/2;z)+L(0;z)
\end{equation}
which is symmetric in $\pm f$, positive and vanishes only at $f=0$. Moreover, 
\begin{equation}\label{lo}
L(0;z) = \psi^\star(D^\dagger \id S/2;z), \qquad L(j;z) = -\frac{1}{2}\,j\cdot D^\dagger \id S(z) + \psi(j;z) +\psi^\star(D^\dagger \id S/2;z)
\end{equation}
which implies that the zero-cost flow must satisfy
\begin{equation}\label{pH2}
   j_z =\partial H(0;z)=\partial\psi^\star(D^\dagger \id S/2;z)
\end{equation}
(If say in one dimension for $j\in \bbR$ we find $f\in \bbR$ for which $\psi(j)+\psi^*(f) = jf$ for Legendre convex pairs $\psi$ and $\psi^*$, then $j=(\psi^*)'(f)$.)
We conclude from \eqref{pH2} that the typical path $\dot z=Dj_z$ has indeed the generalized gradient flow structure \eqref{gger}.

Additonal insight can be gained from adding to \eqref{lo} that
\begin{eqnarray}\label{Hpsistar2}
\psi^\star(f;z) &=& H(f-D^\dagger \id S/2;z)-H(-D^\dagger \id S/2;z)\nonumber\\
H(f;z) &=& \psi^\star(f+D^\dagger \id S/2;z)-\psi^\star(D^\dagger \id S/2;z)
\end{eqnarray}
We see that $\psi$ and $\psi^\star$ are the re-centred symmetric parts of $L$ and $H$ respectively, identical to $L$ and $H$ for $\id S(z)=0$.\\

We now revisit the examples presented in section \ref{grf} in the context of large deviations. 

\subsubsection{Example 1 bis: diffusion limits}\label{ex1b} 

We start with the first example from Section \ref{ex1}, namely the overdamped diffusion. We consider $N$ independent such diffusions. That leads to a quadratic Lagrangian, as encountered in macroscopic fluctuation theory \cite{jona, revjona}, more generally as diffusion limits of interacting particle systems,
\begin{equation}
L(j;\rho)= j\cdot\frac{\chi^{-1}}{4}j-\frac{\id S}{2}\cdot\nabla\chi\nabla\frac{\id S}{2}-\frac{1}{2}j\cdot \nabla \id S
\end{equation}
where $\chi^{-1}$ should be understood as a pseudo-inverse.
The associated Hamiltonian is 
\begin{equation}
H(f;\rho)=f\cdot\chi(f+\nabla\id S)
\end{equation}
It is straightforward to check that $L$ and $H$ verify all the properties discussed above, with functions
\begin{equation}
\psi(j;\rho)=j\cdot\frac{\chi^{-1}}{4}j~~~~~\mathrm{and}~~~~~\psi^\star(f;\rho)=f\cdot\chi f
\end{equation}

\subsubsection{Example 2 bis: jump processes}\label{ex2b} 

The example from Section \ref{ex2} considers a large number $N$ of particles jumping on the graph with vertices (states) $c$ over bonds where $\varphi(c,,c')\neq 0$ with detailed balanced rates $k(c,c')= \varphi(c',c)~\mathrm{e}^{\frac{\beta}{2}(V(c)-V(c'))}$ for $\varphi(c',c)=\varphi(c,c')$. Instead of the Gaussian noise appropriate for a diffusion process, all the currents here have Poissonian statistics, leading to the Lagrangian found in \cite{epl},
\begin{align}
L(j;\rho)=2\sum\limits_{c\sim c'}&\varphi(c',c)\sqrt{\rho(c)\rho(c')}~\lambda\!\left(\frac{j(c,c')}{2\varphi(c',c)\sqrt{\rho(c)\rho(c')}} \right)\\
&+k(c,c')\rho(c)+k(c',c)\rho(c')+\frac{1}{2}j(c,c')\,\log\left(\frac{k(c,c')\,\rho(c)}{k(c',c)\,\rho(c')}\right)\nonumber
\end{align}
with function
 \begin{equation}\label{lambda}
\lambda(j)=j \log\left(j+\sqrt{1+j^2}\right)-\sqrt{1+j^2}
\end{equation}
having Legendre transform $\lambda^\star(f)=\cosh f$. The corresponding Hamiltonian is
\[
H(f;\rho)=2\sum\limits_{c\sim c'}\varphi(c',c)\,\sqrt{\rho(c)\rho(c')}\cosh\!\left(f_{c',c}+\frac{1}{2}\log\frac{k(c,c')\,\rho(c)}{k(c',c)\,\rho(c')}\right)-k(c,c')\rho(c)-k(c',c)\rho(c')
\]
where the sum is taken over every pair of neighboring states, and which can be written more compactly as
\begin{equation}\label{Hjump}
H(f;\rho)=\sum\limits_{c,c'}k(c,c')\rho(c)\left({\mathrm e}^{f_{c',c}}-1\right)
\end{equation}
with $f_{c,c'}=-f_{c',c}$. It is straightforward to check that \eqref{Lpsi} and \eqref{Hpsistar2} hold with 
\begin{equation}
\psi(j;\rho)=2 \varphi(c',c)\sqrt{\rho(c)\rho(c')}~\lambda\!\left(\frac{j}{2 \varphi(c',c)\sqrt{\rho(c)\rho(c')}}\right)+2 \varphi(c',c)\sqrt{\rho(c)\rho(c')}
\end{equation}
and
\begin{equation}\label{jp}
\psi^\star(f;\rho)=2\sum\limits_{c\sim c'} \varphi(c',c)\sqrt{\rho(c)\rho(c')}\left(\cosh f_{c,c'}-1\right)
\end{equation}
and with $S$ given by \eqref{Sjump}, and $\id S$ in the $\cosh$ in \eqref{Hjump}.

\subsubsection{Example 3 bis: Ehrenfest model}\label{ex3b}

The Ehrenfest model in example \ref{ex3} is obtained as a scaling limit of a jump process with jumps of the form $x\rightarrow x\pm \frac{2}{N}$ and associated rates $k^\pm(x)=N\frac{1\mp x}{2}e^{-\frac{N\beta}{2} \big(V(x\pm 2/N)-V(x)\big)}$.

We can immediately write the Lagrangian as a special case from the previous section, with a distribution $\rho(x)=\delta(x)$ localised at $x$. The current $j$ as used there stands for the number of jumps per unit time between $x$ and $x\rightarrow x+ \frac{2}{N}$, so that $\dot{x}=2N^{-1}j$. The Lagrangian becomes
\begin{equation}
L(\dot{x};x)=\sqrt{1-x^2}\lambda\!\left(\frac{\dot{x}}{2\sqrt{1-x^2}}\right)+\sqrt{1-x^2}\cosh (-\beta w'(x)) -\dot{x}\beta w'(x)
\end{equation}
with the same free energy $w(x)= V(x)-\beta^{-1}\eta(x)$ as before and function $\lambda$ defined in \eqref{lambda}. The Hamiltonian becomes
\begin{align}
H(f;x)&=\frac{1-x}{2} e^{-\beta V'(x)} \left(e^{2f} - 1\right) + \frac{1+x}{2} e^{\beta V'(x)} \left(e^{-2f} - 1\right)\nonumber\\
&=\sqrt{1-x^2}\cosh (2f-\beta w'(x)) -\sqrt{1-x^2} \cosh (-\beta w'(x))
\end{align}
Equations \eqref{Lpsi} and \eqref{Hpsistar2} are verified with 
\begin{equation}
\psi(\dot{x};x)=\sqrt{1-x^2}\lambda\!\left(\frac{\dot{x}}{2\sqrt{1-x^2}}\right)+\sqrt{1-x^2},\quad\psi^\star(f;x)=\sqrt{1-x^2}(\cosh (2f) - 1)
\end{equation}

\subsection{From generalized detailed balance to GENERIC}\label{gensec}

An essential ingredient in the previous analysis and examples was the (standard) condition of detailed balance \eqref{detb}.  Looking back at the Examples \ref{ex4}--\ref{ex6} for GENERIC flow we see that it is important to introduce momentum or velocity degrees of freedom to naturally give rise to the additional Hamiltonian flow.   While such systems remain generalized reversible --- one needs to flip the velocities when reversing the spatial trajectory --- the condition \eqref{detb} must be revisited.\\

Let us start with the constraint \eqref{stud} on \eqref{prob}.  We no longer require \eqref{detb} but we suppose indeed that the Lagrangian satisfies
\begin{equation}\label{tfo}
L(J-j;z) - L(J+j;z) = j \cdot D^\dagger \id S
\end{equation}
for some function $J(z)$, to be derived from the Lagrangian as well. The reasoning of section \ref{grgr} can then be repeated by replacing $L$ with $\tilde{L}(j;z) = L(J+j;z)$. The new Lagrangian $\tilde L$ inherits the relevant properties of $L$: non-negativity, convexity in its first argument, and unique minimization.  The Hamiltonian  $H$ is obtained as before in \eqref{defH}, but the generalized detailed balance condition \eqref{tfo} leads to a modified version of \eqref{dbH}, 
\begin{equation}\label{dbH2}
H(-D^\dagger \id S/2-f;z)-H(-D^\dagger \id S/2+f;z)=-2f\cdot J
\end{equation}
making the duality between $L$ and $H$ more symmetric than before.

The functions $\psi$ and $\psi^\star$ are still defined as the re-centered symmetric parts of $L$ and $H$, and are still Legendre transforms of one another. 
They verify
\begin{align} 
L(j;z)&=\psi(j-J;z)+\psi^\star(D^\dagger \id S/2;z)-\frac{1}{2}(j-J)\cdot D^\dagger \id S \label{LgenGEN}\\
H(f;z)&=\psi^\star(f+D^\dagger \id S/2;z)-\psi^\star(D^\dagger \id S/2;z)+f\cdot J\label{HgenGEN}
\end{align}
and the typical current $j_z$ or zero-cost flow is given by
\begin{equation}\label{fro}
j_z = \partial H(0;z)=J + \partial\psi^\star(D^\dagger \id S/2;z)
\end{equation}
Adding the orthogonality condition $J\cdot D^\dagger \id S=0$ leads naturally to the nonlinear version of pre-GENERIC, with $\dot z$ given by \eqref{preGENnlin} for $\cal D =D$,
\begin{equation}
\dot z=D J+D\partial\psi^\star(D^\dagger \id S/2;z)
\end{equation}

We see next how that structure arises from the large deviations of the examples in Section \ref{GENpreGEN}.

\subsubsection{Example 4 bis: underdamped diffusions}\label{ex4b}

Concerning the Vlasov-Fokker-Planck equation of Section \ref{ex4} we can use the fluctuation theory in \cite{DPZ13}. We have a quadratic Lagrangian,
\begin{equation}\label{ul}
L(j;\rho)=(j-J)\cdot\frac{\chi^{-1}}{4}(j-J)-\frac{\id S}{2}\cdot\nabla\chi\nabla\frac{\id S}{2}-\frac{1}{2}j\cdot \nabla \id S
\end{equation}
with 
 Hamiltonian current,
\begin{equation}
J=\begin{bmatrix}\rho\frac{p}{m} \\ -\rho(\nabla_q V + \nabla_q\Phi\star\rho) \end{bmatrix}
\end{equation}
The associated Hamiltonian to \eqref{ul} is
\begin{equation}
H(f;\rho)=f\cdot\chi(f+\nabla\id S)+f\cdot J
\end{equation}
We thus recover the structure of \eqref{LgenGEN} and \eqref{HgenGEN}, with the same functions $\psi$ and $\psi^\star$ as for the detailed balanced overdamped case in Section \ref{ex1b}:
\begin{equation}
\psi(j;\rho)=j\cdot\frac{\chi^{-1}}{4}j,\qquad\psi^\star(f;\rho)=f\cdot\chi f
\end{equation}
from which the equation \eqref{grafi} derives its structure.

\subsubsection{Example 5 bis: Andersen thermostat}\label{ex5b}

We now consider $N$ particles evolving according to the thermostat dynamics of Section \ref{ex5}, i.e., according to \eqref{rhodotAnderson}.  The only origin of randomness is the momentum resetting in the jump-part, while the Hamiltonian flow continues.  We are therefore in the situation of \eqref{str} where $J=\rho K\,\nabla \frac{\delta{\cal H}[\rho]}{\delta \rho(r)}$ is the Hamiltonian current corresponding to 
\begin{equation}
\begin{bmatrix}\dot q \\ \dot p \end{bmatrix}=K\,\nabla \frac{\delta{\cal H}[\rho]}{\delta \rho(r)}=\begin{bmatrix}\frac{p}{m} \\ -\nabla_q V(q) \end{bmatrix}
\end{equation}
For \eqref{prob} we only need to estimate the probability of Poisson-distributed jumps but that is the same as for the Markov jump processes in Section \ref{ex2b}.  We thus have for the current $j$ corresponding to the momentum jumps the expressions
\begin{equation}
\psi(j;\rho)=2 \varphi\int  \frac{e^{-\frac{p^2+p'^2}{4m}}}{\sqrt{2\pi m}} \sqrt{\rho(q,p)\rho(q,p')}\left(\lambda\left(\frac{\sqrt{2\pi m}e^{\frac{p^2+p'^2}{4m}}j_{q,p;q,p'}}{2 \varphi  \sqrt{\rho(q,p)\rho(q,p')}}\right)+1\right)\,\id p \id p'
\end{equation}
and
\begin{equation}
\psi^\star(f;\rho)= 2 \varphi\int \frac{e^{-\frac{p^2+p'^2}{4m}}}{2\sqrt{\pi m}} \sqrt{\rho(q,p)\rho(q,p')}\left(\cosh f_{q,p;q,p'} -1\right)\,\id p \id p'
\end{equation}
as before in \eqref{jp}.

\subsubsection{Example 6 bis: nonlinear friction}\label{ex6b}

In the example from Section \ref{ex6} we have $\cal D = D = I$ and we are in the case with constraint \eqref{stud} in \eqref{prob}.   As in Section \ref{ex3b}, we can directly write the Lagrangian for the dynamics defined in \eqref{BackGenE6}, with a Hamiltonian flow and jumps of the form $p\rightarrow p\pm N^{-1}$ at fixed $q$, with rates $k^\pm(q,p)=N \varphi~e^{-N\frac{(p\pm N^{-1})^2-p^2}{4m}}$ :
\begin{equation}
L(\dot q,\dot p;q,p)=2\varphi~ \lambda\!\left(\frac{\dot p +\nabla_q V(q)}{2\varphi} \right)+2\varphi~\cosh\!\left(\frac{p}{2m}\right)+\dot p \frac{p}{2m}
\end{equation}
with constraint $\dot q=\frac{p}{m}$ and $\lambda$ as in \eqref{lambda}.  The associated Hamiltonian is
\begin{equation}
H(f_q,f_p;q,p)=2\varphi~\cosh\left(f_p+\frac{p}{2m}\right)-2\varphi~\cosh\!\left(\frac{p}{2m}\right)+f_q~\frac{p}{m}-f_p\nabla_q V(q)
\end{equation}
These can be decomposed along equations \eqref{LgenGEN} and \eqref{HgenGEN}, with functions
\begin{equation}
\psi(\dot q,\dot p;q,p)=2\varphi~\lambda\!\left(\frac{\dot p}{2\varphi} \right)+2\varphi~~~~\mathrm{and}~~~~\psi^\star(f_q,f_p;q,p)=2\varphi\left(\cosh f_p-1\right)
\end{equation}
giving the required structure of \eqref{nlf}.

\section{Conclusions}

We have followed the tradition of establishing structural facts about the macroscopic evolution towards equilibrium from symmetries in the dynamical fluctuations.  Our case was that of identifying the appropriate fluctuation symmetry that yields (nonlinear) gradient flow and (generalized) GENERIC.  The latter are equations where a gradient flow is added to the Hamiltonian part in the evolution.  Such evolutions occur when apart from underdamped translational motion of particles there is also a reaction mechanism, such as resetting of momentum, friction or dissipative Langevin forces.  We have given the dynamical fluctuation functions from which the structure of the zero-cost flow can be derived.  An essential ingredient here is the decomposition of the Lagrangian in the path-space action in its entropic and frenetic contributions \cite{frenb}.  The entropic part is antisymmetric under time-reversal, possibly recentred by the Hamiltonian flow and is given in terms of negative free energies.  The frenetic part is the time-symmetric counterpart and contains the nonlinear mobility governing the speed at which the free energy is going to decrease.\\\

\noindent {\bf Acknowledgment}  R.K. was supported by the Deutsche Forschungsgemeinschaft (DFG)
via RTG 2131 High-dimensional Phenomena in Probability Fluctuations and Discontinuity.  A.L. was supported by the Interuniversity Attraction Pole - Phase VII/18 (Dynamics, Geometry and Statistical Physics) at the KU Leuven and the AFR PDR 2014-2 Grant No. 9202381 at the University of Luxembourg. M.A.P. was supported by NWO VICI grant 639.033.008.

\end{document}